\documentclass[aip, preprint, pop]{revtex4-1} 
\usepackage{times}		
\usepackage{booktabs}
\usepackage{placeins}

\usepackage{amssymb,amstext,amsmath, epsf}
\usepackage{graphicx,graphics,epsfig,wrapfig}
\usepackage{listings}
\usepackage{float}
\usepackage{esint}

\usepackage{color}
\usepackage[english]{babel}
\usepackage[T1]{fontenc}
\usepackage[utf8]{inputenc}
\usepackage[font=small,labelfont=bf]{caption}
\usepackage{mwe}    
\usepackage{subcaption}
\usepackage[normalem]{ulem}

\usepackage[normalem]{ulem}

\begin{document}

\title{Hybrid Paricle-in-Cell Simulations of Electromagnetic Coupling and Waves From Streaming Burst Debris}

\author{Brett D. Keenan}
\email{keenan@lanl.gov}
\affiliation{Los Alamos National Laboratory, Los Alamos, NM 87545, USA}
\author{Ari Le}
\affiliation{Los Alamos National Laboratory, Los Alamos, NM 87545, USA}
\author{Dan Winske}
\affiliation{Los Alamos National Laboratory, Los Alamos, NM 87545, USA}
\author{Adam Stanier}
\affiliation{Los Alamos National Laboratory, Los Alamos, NM 87545, USA}
\author{Blake Wetherton}
\affiliation{Los Alamos National Laboratory, Los Alamos, NM 87545, USA}
\author{Misa Cowee}
\affiliation{Los Alamos National Laboratory, Los Alamos, NM 87545, USA}
\author{Fan Guo}
\affiliation{Los Alamos National Laboratory, Los Alamos, NM 87545, USA}

\begin{abstract}
Various systems can be modeled as a point-like explosion of ionized debris into a magnetized, collisionless background plasma -- including astrophysical examples, active experiments in space, and laser-driven laboratory experiments. Debris streaming from the explosion parallel to the magnetic field may drive multiple resonant and non-resonant ion-ion beam instabilities, some of which can efficiently couple the debris energy to the background and may even support the formation of shocks. We present a large-scale hybrid (kinetic ions + fluid electrons) particle-in-cell (PIC) simulation, extending hundreds of ion inertial lengths from a 3-D explosion, that resolves these instabilities. We show that the character of these instabilities differs notably from the 1-D equivalent by the presence of unique transverse structure. Additional 2-D simulations explore how the debris beam length, width, density, and speed affect debris-background coupling, with implications for the generation of quasi-parallel shocks.
\end{abstract}

\maketitle

\section{Introduction}
\label{s:intro}

Explosions of ionized material into a background of magnetized plasma occur in a variety of systems. On very large astrophysical scales, examples include the early stages of supernova remnants.\cite{spicer:1990} Space physics examples include active experiments releasing barium or other chemicals into the magnetosphere \cite{haerendel:1986,johnson:1992} and historical high-altitude nuclear tests.\cite{dyal:2006} Similar physics is explored on smaller scales in the laboratory using lasers to ablate material off solid targets.\cite{borovsky:1984,zakharov:2003,constantin:2009,niemann:2013} 
\newline
\indent
In the above examples, the plasmas are nearly collisionless, and coupling between the fast debris ions and the surrounding magnetized background is mediated by the electromagnetic fields and waves. In the direction perpendicular to the background magnetic field, the coupling mainly occurs via the ``Larmor'' mechanism,\cite{golubev:1979,bashurin:1983,bondarenko:2017} whereby background ions are accelerated in the $\bf{E\times B}$ field induced by the bulk motion of the debris. This generates a diamagnetic cavity \cite{winske:2019} by expelling background magnetic field and plasma, and it may launch a perpendicular shock in the background -- as observed in experiments at UCLA. \cite{schaeffer:2012,schaeffer:2014,clark:2013}
\newline
\indent
A fraction of the debris streams away from the explosion parallel to the background magnetic field. Near parallel angles to the magnetic field, any coupling or shocks that may appear must be mediated by plasma instabilities. Many such beam-driven instabilities are possible. \cite{gary:1984,gary:1991,weidl:2019} The right-hand resonant ion-ion beam instability (RHI) has garnered much attention, and it has recently been observed in the laboratory.\cite{heuer:2020} In addition, a non-resonant right-hand ion-ion beam instability (NRI) is also possible. The joint coupling of the RHI and NRI (along with a left-hand resonant ion-ion beam instability) can lead to the formation of quasi-parallel shocks.\cite{weidl:2016} The NRI and associated nonlinear structures have, for example, been observed by the Magnetospheric Multiscale Mission (MMS) at the Earth's quasi-parallel bow shock and are accurately reproduced by fully kinetic 1-D simulations.\cite{chen:2021} The NRI may also be within the reach of existing -- or slightly modified -- experimental platforms.\cite{weidl:2016,heuer:2020par}
\newline
\indent
Modeling the coupling of debris ions from an explosion to the background is challenging because it is difficult to resolve the relatively small fraction\cite{le:2021} of debris that escapes parallel to the magnetic field and because the coupling occurs over long length scales of hundreds of ion inertial lengths. Retaining ion kinetics in such a large system is possible when the electrons may be treated as a neutralizing background, requiring only a generalized Ohm's law to couple the electron fluid to the kinetic ions. Such hybrid approaches,\cite{lipatov:2002,hewett:2011,winske:2007,clark:2013} where a multi-ion plasma is evolved with a particle-in-cell (PIC) code have been verified by comparisons with fully kinetic simulations and validated with laboratory experiments. Using a hybrid version of the Los Alamos PIC code VPIC,\cite{bowers:2008, le:2021} we explore the growth and development of ion-ion beam instabilities following an idealized point-like 3-D explosion in a magnetized plasma background. 
\newline
\indent
We apply a 1-D spectral analysis, verified with simplified 1-D simulations in Appendix \ref{sec:1-D_verify}, to find evidence for the RHI and NRI in the 3-D simulation. The coupling between the ion beams and the background plasma is also consistent with 1-D simulations. The analysis, however, is complicated by the finite width and length of the debris beams escaping from the explosion, as well as non-uniformities that naturally develop. We use simplified 2-D simulations with beams of finite width and length to explore these limiting factors on growth rates and wavelengths in a more controlled setting. The finite width of the beam in the transverse direction modifies the mode structure of the electromagnetic fields of the beam-driven instabilities. While the growth rates of the ion-ion beam instabilities depend mainly on the beam density and velocity, the overall coupling between the beam and the background depends also on the length of the beam,\cite{onsager:1991} which sets the length over which the instabilities may grow. The local beam parameters are therefore not sufficient in themselves to determine the level of coupling, or the possibility of quasi-parallel shock formation. Moreover, it is difficult to characterize the beam and background parameters because the most significant coupling occurs after the fluctuations reach order-unity amplitude and perturb the local conditions. While the 1-D linear theory -- as detailed in Appendix \ref{sec:1-D_beam} -- offers very useful intuition, it cannot account for all the details of the beam dynamics.    
\newline
\indent
The paper is organized as follows. Section \ref{sec:3-D} describes a 3-D hybrid VPIC simulation of a point-like blast, and it shows an inhomogeneous beam of parallel-streaming debris with a finite length and width. We show that the finite width of the debris beam implies a certain wave structure in 3-D. Using spectral analysis, Section \ref{sec:hane_instab} shows that ion beam instabilities drive the production of waves and electromagnetic coupling in the 3-D simulation. Next, Section \ref{sec:finitebeam} addresses the impact of the debris beam's finite length and width on the debris/background coupling, as well as the implications for quasi-parallel shock formation. We will also show that some of the background ions are noticeably accelerated via this coupling. Finally, we conclude in Section \ref{sec:disc}.

\section{Fast Ions and Waves in a 3-D Explosion Simulation}
\label{sec:3-D} 

We perform a 3-D calculation using the Hybrid-VPIC code of an astrophysical explosion similar to the 2-D hybrid simulations of Winske \& Gary.\cite{winske:2007} The electron model in Hybrid-VPIC reduces to a generalized Ohm's law for the electric field without electron inertia:
\begin{equation}
{\bf{E}} = -{\bf{u_i\times B}} - \frac{1}{en_e}\nabla p_e +  \frac{1}{en_e}{\bf{J \times B}} - \eta{\bf{J}} + \eta_H \nabla^2{\bf{J}}
\end{equation}
where $p_e$ and $n_e$ are the electron pressure and density, respectively; electrons are treated as an isothermal fluid; ${\bf B}$ is the magnetic field; quasi-neutrality is enforced by setting the electron number density, $n_e=\sum_sZ_sn_s$ (where the sum is over ion species, $s$, and $Z_s$ is the ion charge state); the ion velocity appearing in Ohm's law is charge-weighted; i.e., ${\bf{u_i}}=\sum_s Z_sn_s{\bf{u_s}}/n_e$; and the current density, ${\bf J}$ is taken in the low-frequency approximation: $\mu_0{\bf{J}} = \nabla\times{\bf{B}}$. The normalized resistivity $\eta/(B_0/n_be)$ and hyper-resistivity $\eta_H/(B_0/n_bed_i^2)$ are in the range of $[0.001, 0.005]$, which aids with numerical stability and agrees with previous comparisons of fully kinetic and hybrid models.\cite{le:2016,stanier:2017,wetherton:2019} We use nearest-grid-point particle weights in these simulations to avoid numerical dispersion at shocks, which can occur with higher-order particle schemes in hybrid PIC algorithms.\cite{stanier:2020}

The initial conditions contain a uniform background plasma composed of electrons and ions of mass $m_i$ and charge $Z_i$ of number density $n_i=n_0$ in a uniform magnetic field, $B_x=B_0$. The initial electron and ion temperature $T_e=T_i$ is relatively low, and it gives a background plasma beta of $\beta = 2\mu_0n_0(T_e+T_i)/B_0^2 = 0.1$. The simulation domain is $L_x \times L_y \times L_z = 1296 \times 540 \times 540$ cells with a uniform grid spacing of $1~d_i$ (where $d_i$ is the ion inertial length  $d_i = \sqrt{\epsilon_0m_ic^2/Z_i^2e^2n_i}$ based on the background plasma density). The time step is $\delta t \times \omega_{ci} = 0.005$ (where $\omega_{ci} = Z_ieB_0/m_i$ is the background ion cyclotron frequency based on the background field), and the background plasma is sampled with $\sim 150$ particles per cell ($\sim$56 billion total). 

A cloud of debris ions of mass $m_d/m_i=3$ is initialized at the center of the domain with a spatial profile given by $n_d = n_*\exp(-r^2/\Lambda_*^2)$, where the scale length $\Lambda_*=4~d_i$ and $n_*$ is chosen so that the total debris mass is equal to the background plasma mass contained in a sphere with radius (the equal mass radius \cite{winske:2007}) of $R_m=40~d_i$. The debris ion velocity distribution is a drifting Maxwellian with a radial explosion drift speed $u_d$ characterized by an Alfv\'{e}n Mach number of $M_A = u_d/v_A=5$ (where $v_A = B_0/\sqrt{\mu_0n_0m_i}$) and a small thermal velocity spread of $v_{thd} = \sqrt{T_d/m_d} = 0.1u_d$. This corresponds to a directed debris gyroradius of $\rho_d = u_d/\Omega_{cd} = 15~d_i$. The debris ions are sampled with $\sim$ 560 million total numerical particles. These parameters, with a ratio of $\rho_d/R_m = 0.375$, were chosen to clearly reproduce strong debris ion to background coupling, and were based on previous simulations and laboratory experimental parameters.\cite{winske:2007,le:2021}

\begin{figure}
	\centering
    \includegraphics[width=1.0\linewidth]{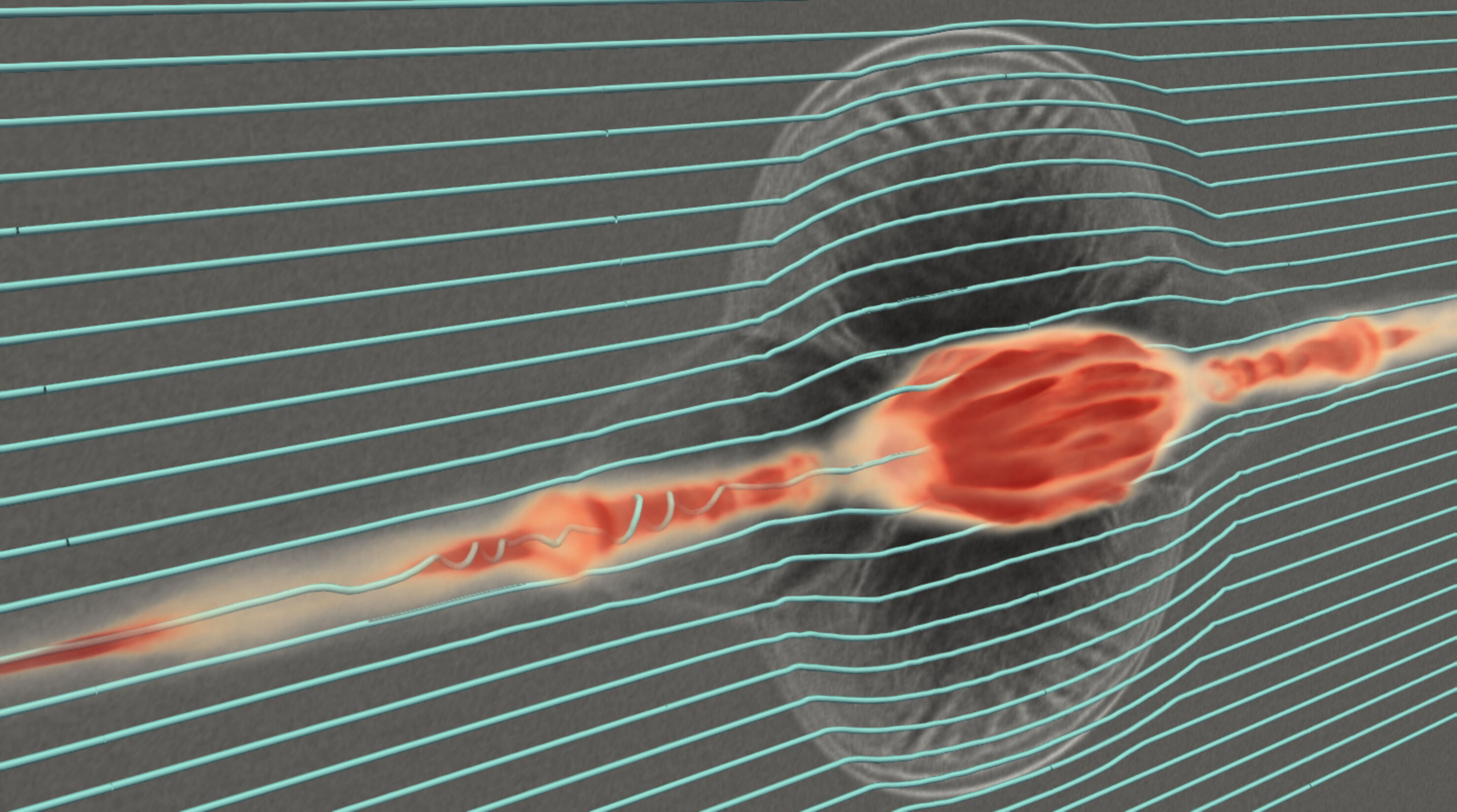}
	\caption{Volume rendering of the debris ion density at $t \times \omega_{ci}=62$ with a slice of the background density (grayscale) and sample magnetic field lines (cyan curves). Fast beams of mainly magnetic field-aligned debris ions drive ion-ion beam instabilities as they stream away from the explosion. }
	\label{fig:3-D}
\end{figure}

To help visualize the overall interaction and dynamics, Figure~\ref{fig:3-D} shows a volume rendering of the debris density $n_d$ in red at time $t \times \omega_{ci}=62$. At this stage, the diamagnetic cavity \cite{winske:2019} has already begun collapsing (the cavity reaches its peak size of $\sim R_m = 40~d_i$ at $t \times \omega_{ci}\sim 25$).  Similar to what was observed in 2-D simulations,\cite{le:2021} there is a central region of high density consisting of debris ions that remain inside the collapsing cavity, and a shock wave propagates out across the magnetic field (as indicated by the outward bending field lines). In addition, there are beams of fast debris ions streaming out through circular ``holes'' at either end of the cavity, and which propagate nearly parallel to the ambient magnetic field in the $x$ direction. The beam speeds range from $\sim2~v_A$ near the cavity to a little over the initial expansion speed of $5~v_A$ at the leading edge. The rotation of the magnetic field lines associated with these beams suggests a helical magnetic field wave structure; which is consistent with the waves generated by the RHI and NRI.\cite{winske:1984}

\begin{figure}
	\centering
    \includegraphics[width=0.8\linewidth]{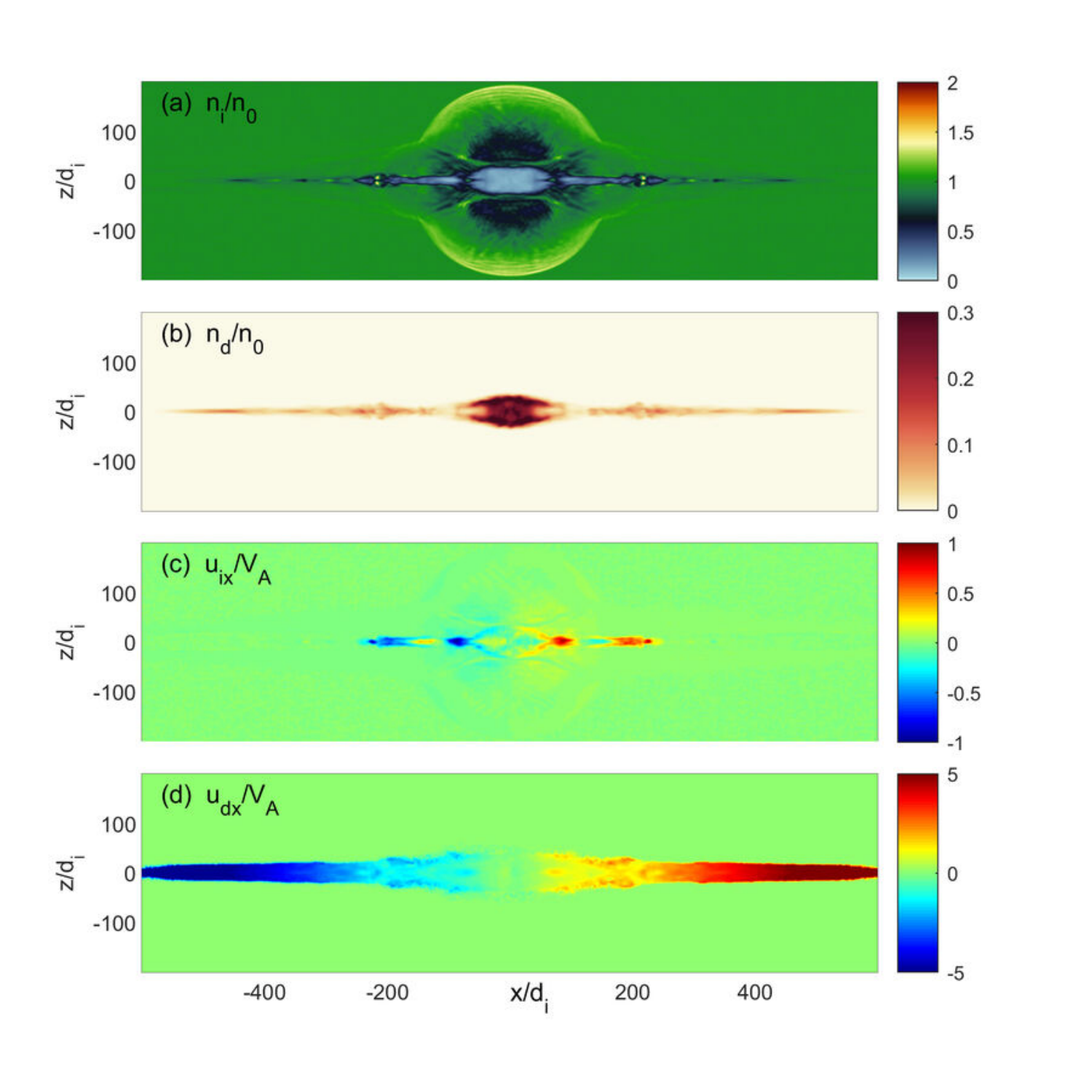}
	\caption{Slices of data from the 3-D simulation at time $t \times \omega_{ci}=90$. Plotted are (a) the background plasma density $n_i$ showing an outgoing shock, (b) the debris ion density $n_d$ including parallel-streaming beams, (c) the background bulk velocity component $u_{ix}$, and (d) the debris velocity $u_{dx}$.}
	\label{fig:3-Dslice}
\end{figure}

Plotted in Fig.~\ref{fig:3-Dslice} are 2-D slices from time $t \times \omega_{ci}=90$ from the center of the 3-D simulation of Fig.~\ref{fig:3-D}. The background ion density in Fig.~\ref{fig:3-Dslice}(a) shows the remaining diamagnetic cavity at the center,\cite{winske:2007,winske:2019} where background magnetic field and plasma has been expelled by the explosion. A shock front propagating primarily in the $z$ direction perpendicular to the background magnetic field is also clearly visible. The debris ion density is plotted in Fig.~\ref{fig:3-Dslice}(b). The majority of the debris ions have been slowed by their Larmor interaction with the surrounding magnetic field and background density,\cite{bashurin:1983} and they remain within the collapsed diamagnetic cavity.

Parallel to the magnetic field, however, coupling to the background is relatively weak. As a result, a portion of the debris streams out along the magnetic field as a fast-moving beam. The bulk debris drift speed in $x$ is plotted in Fig.~\ref{fig:3-Dslice}(d). Note that the debris ion velocity distribution in Fig. 2(d) appears significantly wider in the transverse direction (and longer in the x-direction) than that of the debris ion density in Fig. 2(b) because the velocity distribution is not density-weighted. As a measure of the fraction of the total debris contained within these fast beams, we find that $\sim9\%$ of the debris density is contained in regions where the bulk debris drift satisfies $u_{dx} > 0.9u_d = 4.5v_A$. This fraction of the debris that streams parallel to the magnetic field depends on the total quantity of debris and the explosion speed, as shown in earlier 2-D simulations over a wide range of initial conditions.\cite{le:2021} 

\begin{figure}
	\centering
    \includegraphics[width=1\linewidth]{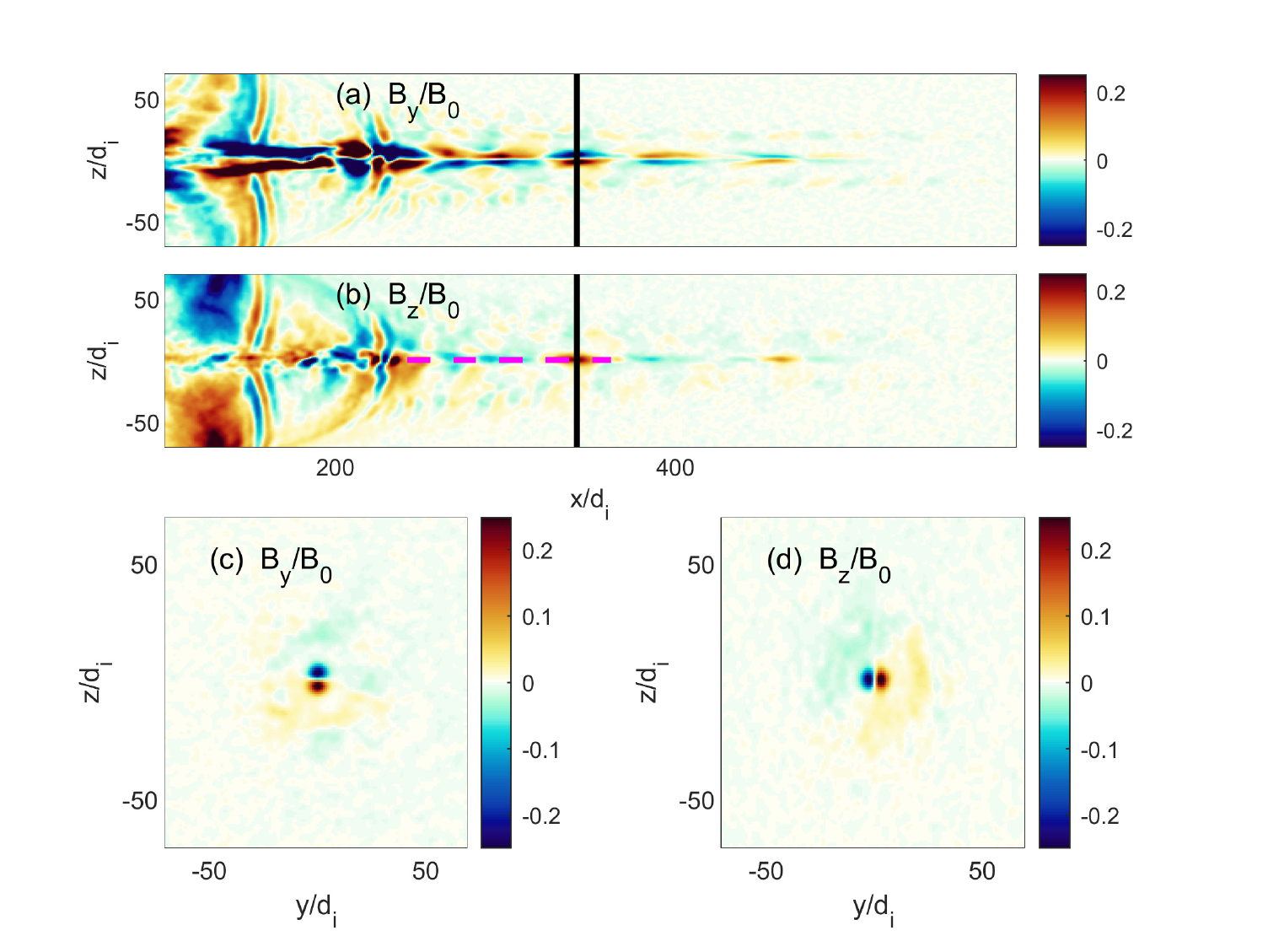}
	\caption{Slices of magnetic field components $B_y$ and $B_z$ in two planes from the 3-D simulation at time $t \times \omega_{ci}=90$ showing waves generated by the parallel-streaming debris ions. Plotted are (a-b) zoomed-in slices in the $x-z$ plane from the center of the simulation (as in Fig.~\ref{fig:3-Dslice}) and (c-d) slices in the $y-z$ plane at the $x$ location marked by the vertical lines in (a-b). The mode structure is qualitatively similar to Eq.~\ref{eq:modelb}. The magenta dashed line in (b) indicates the cut used for the 1-D Fourier analysis of Sec.~\ref{sec:hane_instab}}
	\label{fig:3-Dxslice}
\end{figure}

The parallel-streaming ions interact with the background through collisionless electromagnetic ion-ion beam instabilities.\cite{gary:1991} The fluctuations are evident in Fig.~\ref{fig:3-Dxslice} (a-b), which zooms in on the transverse (to the background field $B_x$ and streaming direction) magnetic field components $B_y$ and $B_z$ at time $t \times \omega_{ci}=90$ as plotted in Fig.~\ref{fig:3-Dslice}. While 1-D linear theory,\cite{gary:1991,weidl:2016} as detailed in Appendix \ref{sec:1-D_beam}, captures basic features of the unstable modes driven by the debris ions streaming from a 3-D explosion, the details of the 3-D electromagnetic fields are different. The width of the beam, which is on the order of the debris ion gyroradius, sets a characteristic transverse length scale. In the case of the example in Fig.~\ref{fig:3-Dslice}, the debris beam radius is  $\sim10~d_i$. This transverse scale length is shorter than the typical longitudinal wavelengths of $\sim 100~d_i$ of the electromagnetic ion-ion beam instabilities. Although this may imply some penalty in terms of the electromagnetic coupling between the beam and the background, the exact details are left to a future study. Suffice to say, the mode structure of any resultant unstable plasma waves will be deviate from the 1-D linear theory prediction because of this disparity in spatial scales.

To compare results to linear theory predictions, we must further qualify the spectral properties of the generated modes. A full eigenmode analysis of a finite 3-D beam, however, is not tractable. Moreover, the mode structure for the finite width beam is not necessarily helical and circularly polarized, as is the case for an infinite width/length beam. Putting aside these issues for the moment, we find that a very simple model of the fluctuating magnetic fields, along with the constraint that $\nabla\cdot{\bf{B}}=0$, reflects the qualitative mode structure of the instabilities observed in the 3-D simulation. Based on intuition from the 1-D theory, we assume the unstable modes have comparable transverse normalized components $\hat{B}_y \equiv B_y/B_0$ and $\hat{B}_z \equiv B_z/B_0$. Given these considerations, we propose that the fluctuating magnetic fields are of the form
\begin{equation}     
\begin{cases}
  \hat{B}_y =F(y,z) \exp(-(y^2+z^2)/2\Lambda_t) \exp(-i\omega t + i k_{||}x) \\
  \hat{B}_z =G(y,z) \exp(-(y^2+z^2)/2\Lambda_t) \exp(-i\omega t + i k_{||}x), \\
\end{cases}
\label{eq:modelb}
\end{equation}
where we assume the mode structure in the transverse directions is modulated by a Gaussian envelope with characteristic transverse dimension $\Lambda_t$. The constraint $\nabla\cdot{\bf{B}}=0$ imposes that
\begin{equation}
    \frac{-y}{\Lambda_t}F + \frac{\partial F}{\partial y} + \frac{-z}{\Lambda_t}G + \frac{\partial G}{\partial z} = 0.
\end{equation}
Note that there is no consistent solution with $F$ and $G$ equal to constants. Instead, the simplest solution is a mode with $F = C_0z $ and $G = -C_0 y$, where $C_0$ is the same constant in each case. This qualitatively matches the magnetic fields driven by the streaming debris ions in the 3-D simulation seen in Figs.~\ref{fig:3-Dxslice}, where $B_y$ is roughly an odd-parity function of $z$ and $B_z$ is roughly an odd function of $y$. Meanwhile, the approximate parallel wave-number $k_{||}\sim0.06/d_i$ (corresponding to a wavelength of $\sim100 d_i$) in the $x$ direction of the modes in Fig.~\ref{fig:3-Dxslice} is very similar to the predictions (described in Section \ref{sec:hane_instab}) from simple 1-D theory. This is perhaps not unexpected because the fluctuations here are dominated by a resonant mode where the parallel wave-number is primarily set by an approximate Alfv\'{e}nic cyclotron resonance $\omega - k_{||}u_d \sim \omega_{cd}$ with $\omega/k_{||}\sim v_A$, where $u_d$ is the parallel debris streaming velocity.

While the instability structure in Fig.~\ref{fig:3-Dxslice} is clearly more complicated than the plane wave solution applied in simple linear theories of homogeneous systems, the linear theory is nevertheless useful for understanding qualitatively (and, in some cases, even quantitatively) what is occurring in the more complicated 3-D simulation. In the following sections, we review the characteristics of the relevant instabilities using linear theory along with simplified 2-D simulations to understand some properties of the complex wave phenomena seen in Figure 3. 

\section{The Signatures of Ion-Ion Beam Instabilities in the 3-D Simulation} 
\label{sec:hane_instab}

Various ion-ion beam instabilities are possible in a magnetized plasma. \cite{gary:1984,weidl:2016,weidl:2019} For cold ion beams moving very fast (several or more times the Alfv\'{e}n speed) along the magnetic field, the dominant instabilities are typically purely parallel ($\bf{k \times B = 0}$) modes. A wider array of unstable modes is possible when the beams are relatively slow (less than or similar to $v_A$) and warm (with thermal speeds comparable to the beam speed or the Alfv\'{e}n speed), and some of the debris-background coupling near the burst region in our 3-D simulation falls into this regime (see Fig.~\ref{fig:3-Dslice}). Interestingly, obliquely propagating modes under these conditions may have even higher growth rates than the purely parallel ion-ion instabilities.\cite{daughton:1998} The growth rates and wavelengths of the oblique modes depend on the beam drift speed, density, temperature, and temperature anisotropy. These plasma conditions vary rapidly in Fig.~\ref{fig:3-Dslice} in the regions near the burst, making a rigorous analysis unfeasible. 

Farther away (beyond $\sim 300~d_i$) from the burst, as shown in Fig.~\ref{fig:3-Dslice}, the debris ion beams are relatively fast and cold. Moreover, in contrast to the complicated coupling at $x \lesssim 300\ d_i$, Fig.\ \ref{fig:3-Dxslice} suggests that the magnetic field in this region has a more regular wavelike structure. Although this mode structure is modified by the transverse beam geometry, as described in Sec.~\ref{sec:3-D}, several characteristics of the wave-like fluctuations along the magnetic field (i.e, the x-axis, or $\parallel$-direction) turn out to agree with a simple 1-D theory of ion-ion beam instabilities (as we show below).

Given the plasma conditions in the vicinity of the region beyond $300\ d_i$, the infinite beam, linear theory\cite{gary:1984} predicts that the purely parallel RHI and NRI modes grow faster than any oblique mode. Given this fact, along with the notable wave characteristics at $x \gtrsim 300\ d_i$, it makes sense to perform a spectral analysis in the parallel or $x$ direction in this region (with $y = z = 0$).
\newline
\indent
\begin{figure}[htbp]
    \centering
        \includegraphics[width=1.0\textwidth]{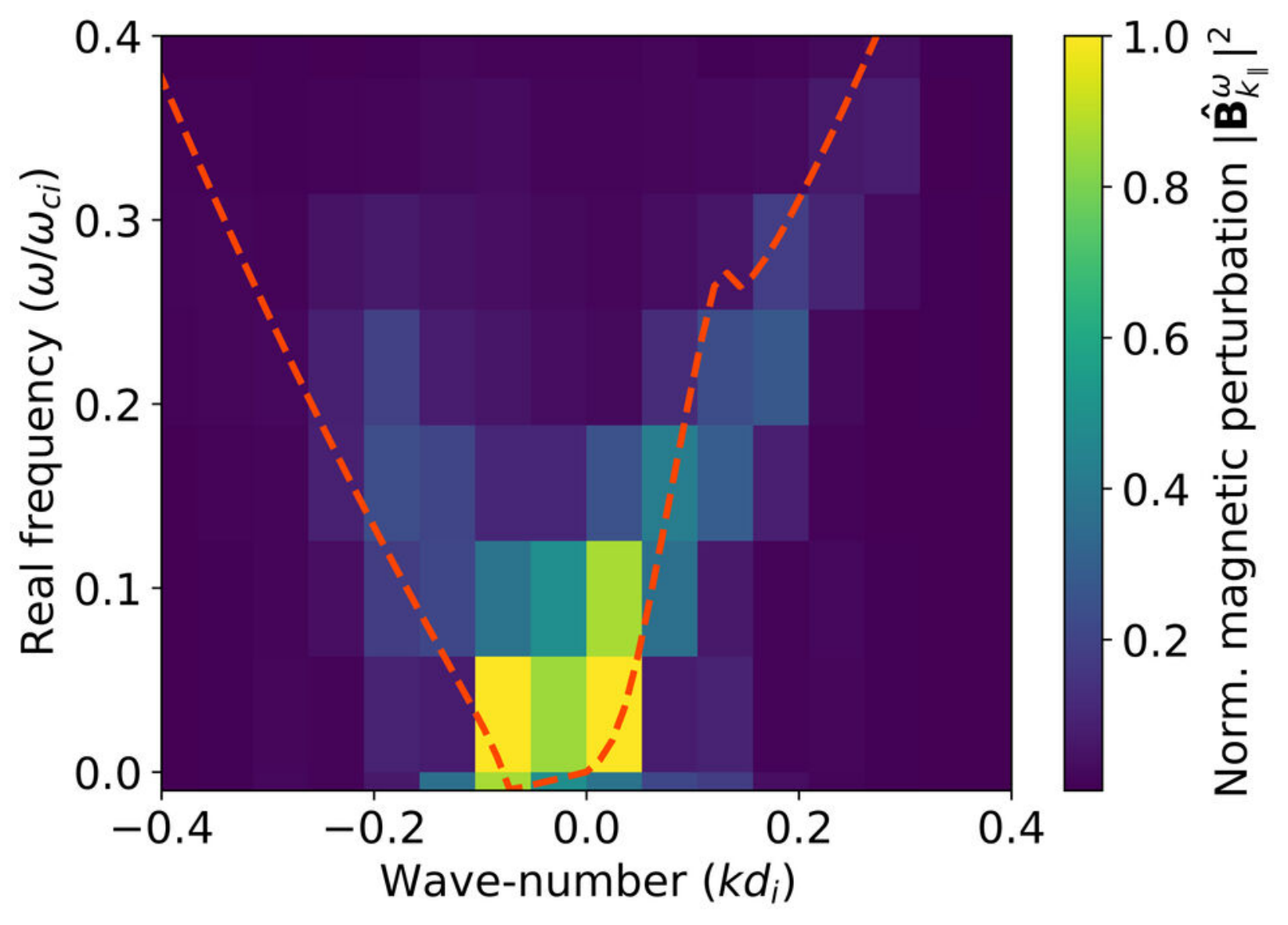}
        \caption{Real frequency vs.\ wave-number from the 3-D hybrid VPIC explosion simulation. The linear theory prediction appears as the red (``dashed'') line. Note that the linear theory prediction for the wave-numbers at maximum growth of the RHI and NRI are $k_\parallel d_i \simeq 0.07$ and $k_\parallel d_i \simeq -0.05$, respectively.}
        \label{omega_vs_k}
\end{figure}
\newline
\indent
We start by considering the real frequency and wavelength of the waves. Using an FFT of the magnetic field on a window in space, $x \in (242-D_i, 362-D_i)$, we see peaks in the fluctuation power vs.\ $k_\parallel$, which are near locations of the maximum instability growth rates predicted by linear theory (not shown), though these peaks are only crudely resolved (as explained below). In Fig.\ \ref{omega_vs_k}, we show a contour plot of the magnetic field FFT power spectrum across a range of $\omega_r$ and \ $k_\parallel$ (once again, using a segment of the x-axis in the 3-D simulation). Unlike the FFT in Appendix \ref{sec:1-D_beam}, the negative $k$ values are distinct from the positive $k$, since the time domain FFT is performed after the k-space transform (i.e., the frequency space transform is on a complex signal, rather than a real one). Overplotted on this figure is a red (``dashed'') curve given by the solution to Eq.\ (\ref{linear_dispersion}), but Doppler-shifted to the lab frame. As before, the theory curve at $k < 0$ represents the NRI, whereas $k > 0$ corresponds to the RHI. Since the ion beam is finite in length and variable in both space and time, the plasma parameters applied to Eq.\ (\ref{linear_dispersion}) are spatially averaged quantities (normalized beam density $n_d/n_e = 0.023$, beam velocity $M_A = 5.66$). Here, we ignore the transverse variation in the beam (i.e., the width), and perform the spatial averages only along x. We see that the bulk of the wave activity (for $k_\parallel > 0$) occurs along the theoretical curve, suggesting that the resonant instability is, indeed, responsible for the growth. Moreover, Fig.\ \ref{omega_vs_k} shows that a peak in the magnetic field occurs around a wave-number close to the expected wave-number for maximum growth of the RHI, $k_\text{max} \simeq 0.07d_i^{-1}$ (see Figure \ref{dispersion_3-D} for a plot of the dispersion relation). On the other hand, there is some indication of the non-resonant mode in the $k_\parallel < 0$ region, but the signal appears weaker. Note that the kink feature at $kd_i \simeq 0.13$ occurs at the point for which the RHI growth rate goes to zero (as shown in Fig.\ \ref{dispersion_3-D}). The waves at higher $k$ then correspond to the standard, parallel Alfv\'{e}n wave branch, with $\omega_r = k_\parallel v_A$. A similar branch transition occurs at $k_\parallel d_i \simeq -0.09$ for the NRI. This may explain the relatively weaker field signal at $kd_i \lesssim -0.09$.
\newline
\indent
\begin{figure}[htbp]
    \centering
        \includegraphics[width=1.0\textwidth]{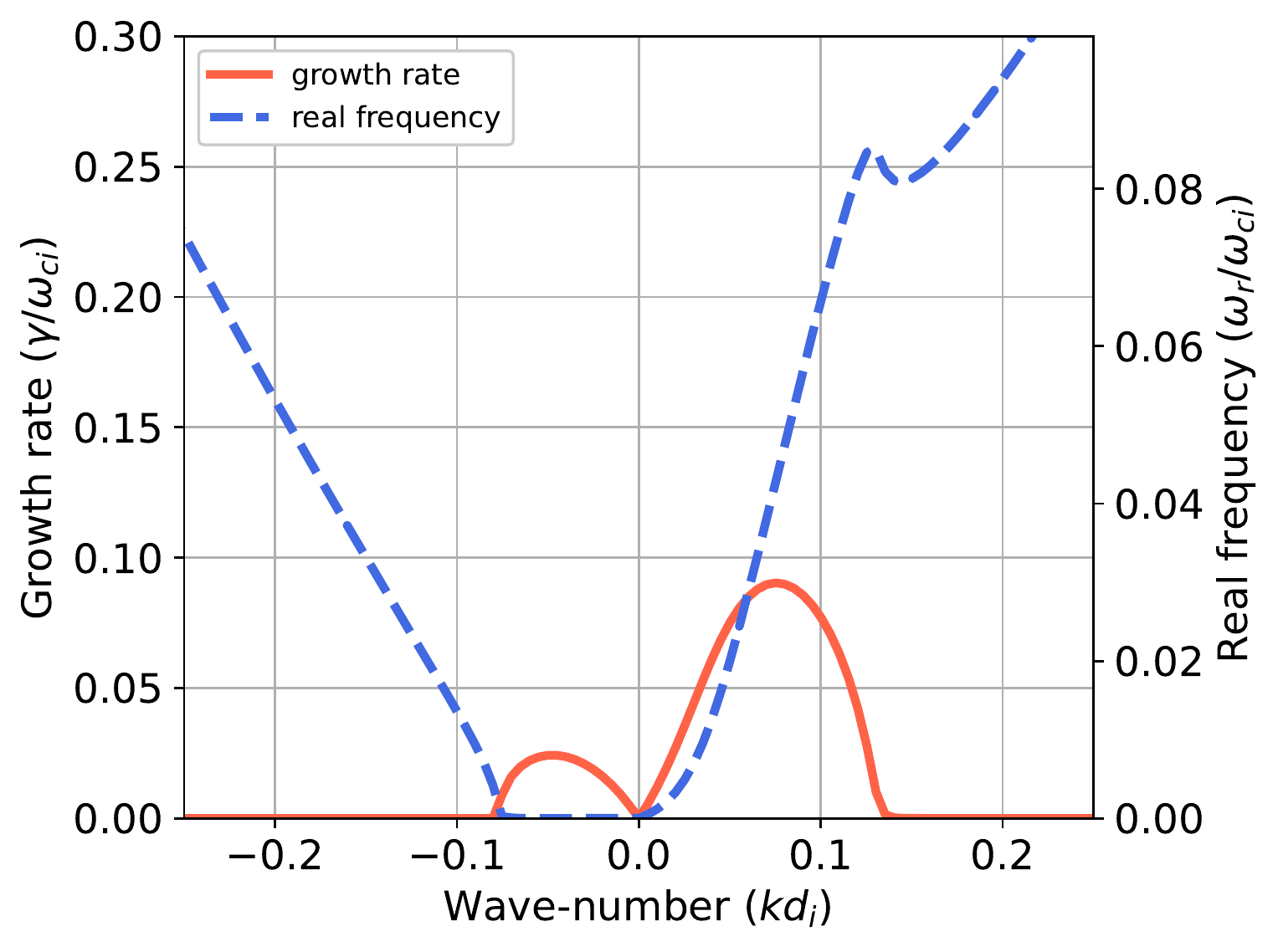}
        \caption{The linear growth rate and real frequency vs.\ wave-number for the RHI and NRI given the plasma conditions on the spatial window: $x \in (242-D_i, 362-D_i)$.}
        \label{dispersion_3-D}
\end{figure}
\newline
\indent
Next, for a fixed real frequency, we plot $|B_{k_\parallel}^\omega|^2$ vs.\ k in Figure \ref{k_bfield}. The real frequency in this plot is $\omega_r = 0.13\omega_{ci}$, which corresponds to the $\omega_r$ at the maximum growth rate for the RHI (given the same average plasma conditions). This plot shows a distinct peak occurs in $|B_{k_\parallel}^\omega|^2$ at $k_\parallel d_i \simeq 0.08$, which is very close to the linear theory prediction: $k_\parallel d_i \simeq  0.07$. Although the NRI is not at its peak growth rate for $\omega_r = 0.13\omega_{ci}$, the linear theory prediction at this frequency, $k d_i = -0.17$ is also close to the Hybrid-VPIC peak at $k d_i \simeq -0.19$. The theoretical maximum growth rates for the RHI and NRI are $0.09\omega_{ci}$ and $0.02\omega_{ci}$, respectively. Note that we show the discrete simulation points in Fig.\ \ref{k_bfield} to emphasize the fact that the maximum growth for both happens near the limit of the resolution afforded by the finite spatial window we do the analysis on. Because of the inhomogeneity of the debris ion beam, it is difficult to select a spatial window with relatively uniform beam parameters that accommodates the full wavelength of both modes. Likewise, it is difficult to compare growth rates to linear theory because the local beam density varies on time scales comparable to the instability growth rates. This is likely why slopes obtained from $|B_{k_\parallel}|^2$ vs.\ time over-predict the instability growth rates.
\begin{figure}
	\includegraphics[width=1.0\linewidth]{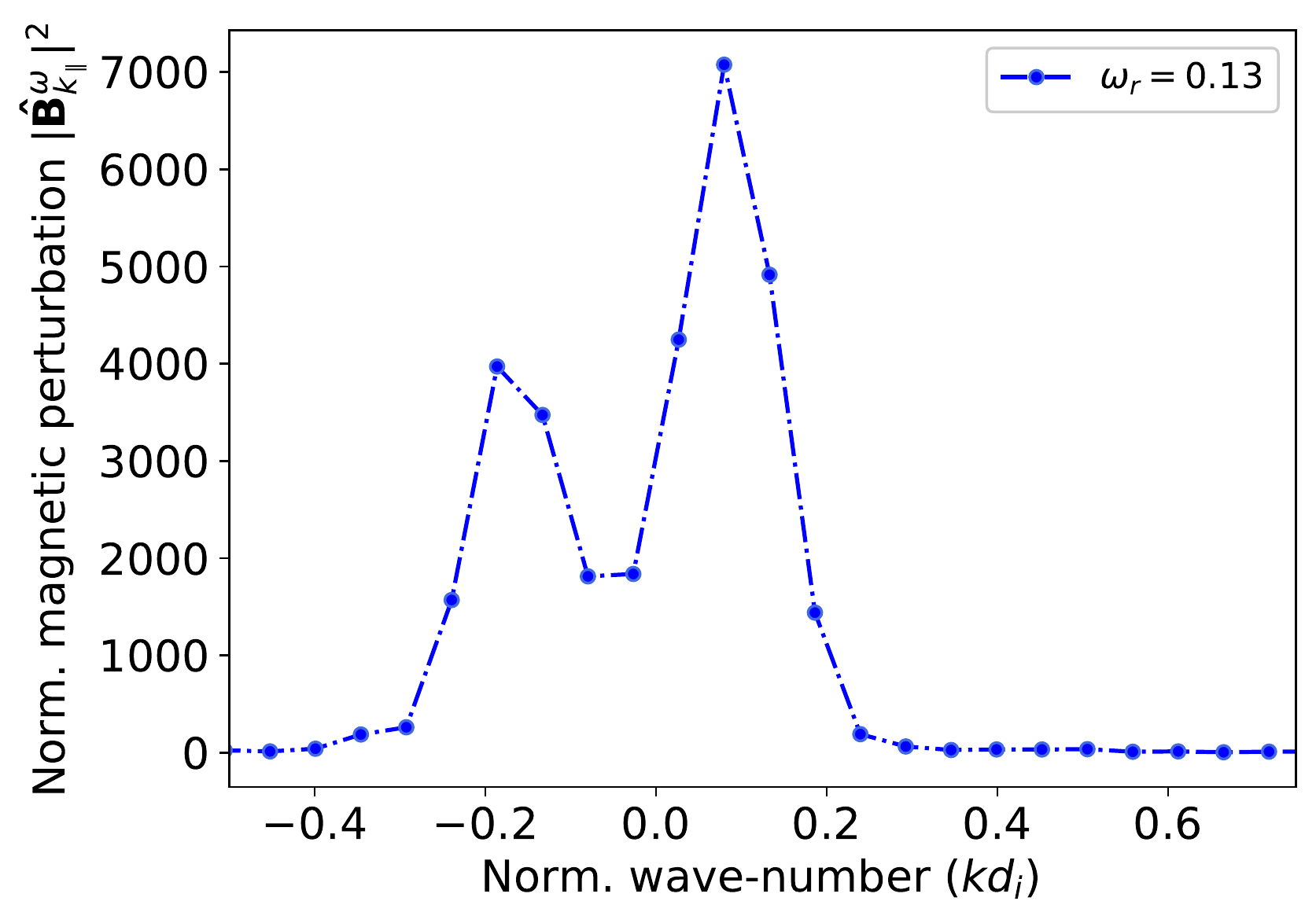}
	\caption{Magnetic perturbation vs.\ the parallel wave-number, along the spatial window: $x \in (242-D_i, 362-D_i)$ at the real frequency for maximum RHI growth ($\omega_r = 0.13\omega_{ci}$). The spectrum peaks at $kd_i \simeq 0.08$, which is close to the theoretical prediction for the RHI; and $k d_i \simeq -0.19$, which roughly corresponds to the linear theory prediction for the NRI.}
	\label{k_bfield}
\end{figure}
Next, we note that we have avoided using the helicity decomposition given by Eq.\ (\ref{hel_decomp}) for this analysis, given the simple fact that the mode structure we observe in the 3-D simulation is far more complicated than the simple circularly polarized modes seen in a 1-D geometry with finite beams. For this reason, we believe that the helicity decomposition may be inappropriate.
\newline
\indent
Our spectral analysis provides evidence for the resonant ion-ion beam instability in the 3-D hybrid explosion simulation and weaker non-resonant modes. Some questions remain, however, given that the ion debris in the 3-D simulation is both spatially and temporally inhomogeneous. The NRI and RHI are local instabilities, and thus how they lead to debris-background coupling in the 3-D simulation needs further clarification. To better understand this, we consider a 2-D hybrid-VPIC simulation of a finite beam in Section \ref{sec:finitebeam}.

\FloatBarrier

\section{2-D Finite Beam Simulations }
\label{sec:finitebeam}

As alluded to above, local instability growth rates alone do not determine whether a beam of finite length and width will strongly couple to the background plasma. In previous 1-D studies, it was found that the beam must have sufficient density and length/duration to strongly couple to a background plasma through a beam-driven instability.\cite{onsager:1991} Here, we consider 2-D simulations that demonstrate the basic beam-background coupling mechanisms explored in the previous sections. Like earlier 1-D studies,\cite{onsager:1991} our 2-D simulations explore the effects of the finite length of the beam. In addition, however, we also include the effects related to the beam's finite transverse width.

The initial conditions for our finite beam simulations contain the same uniform background plasma as in Section \ref{sec:3-D}. A beam of debris ions is initialized at the left edge of the box with a flat top profile of length $\Lambda_x$ in $x$, and a Maxwellian profile in $z$ given by $n_d = n_{d0}\exp(-z^2/2\Lambda_z^2)$. The velocities are sampled from a shifted Maxwellian with bulk flow speed $u_{dx}$ to the right, and a thermal spread given by a temperature $T_d/m_d = (0.1 \times u_{dx})^2$. We use parameters similar to the local conditions where we find the ion-ion beam instability in the 3-D simulation of Sec.~\ref{sec:3-D}: $m_d/m_b=3$, $Z_d/Z_b = 1$, $n_{d0}/n_0 = 0.05$, $u_{dx}/v_A = 7.5$. For these simulations, the domain is $n_x\times n_z=2700\times720$ cells with a grid spacing of $1~d_i$, and the time step is $\delta t \times \omega_{ci}=0.02$. We use $\sim400$ background particles per cell ($\sim780$ million total) and $\sim 16$ million particles to track the debris population.

\begin{figure}
	\centering
    \includegraphics[width=1\linewidth]{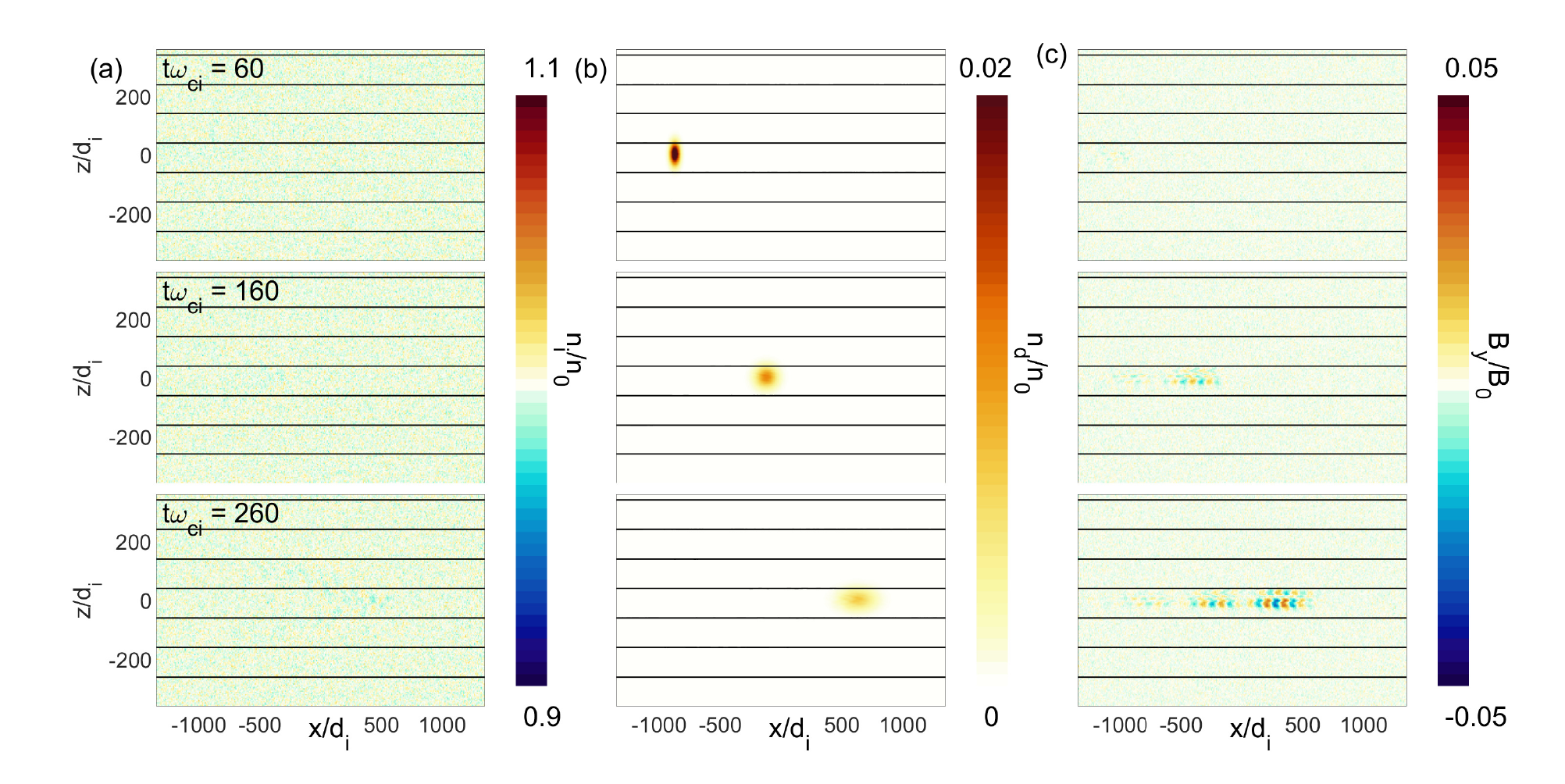}
	\caption{(a) Background density, (b) debris density, and (c) out-of-plane magnetic field component $B_y$, each at three different times, from a 2-D hybrid simulation of a finite debris beam of length $\Lambda_x=50~d_i$, width $\Lambda_z=25~d_i$, peak density of $n_d/n_0=0.03$, and initially moving to the right with an Alfv\'{e}n Mach number of $M_A=7$.}
	\label{fig:finite50}
\end{figure}

The fast ion beam in the simulation of Fig.~\ref{fig:finite50} is initialized with a length of $\Lambda_x=50~d_i$. The local plasma parameters are unstable to both RHI and NRI, with local linear growth rates of $\gamma_{RHI}\sim 0.12\omega_{ci}$ and $\gamma_{NRI}\sim 0.15\omega_{ci}$. Because the finite beam is relatively short, however, the unstable modes develop only over the limited time $\tau\sim\Lambda_x/v_{dx}\sim7/\omega_{ci}$. This is insufficient time for the modes to reach a saturated non-linear state, and the instabilities manifest as a weak wave train following the finite ion beam. It is noteworthy that this situation has been observed in a laser-plasma lab experiment.\cite{heuer:2020par}

\begin{figure}
	\centering
    \includegraphics[width=1\linewidth]{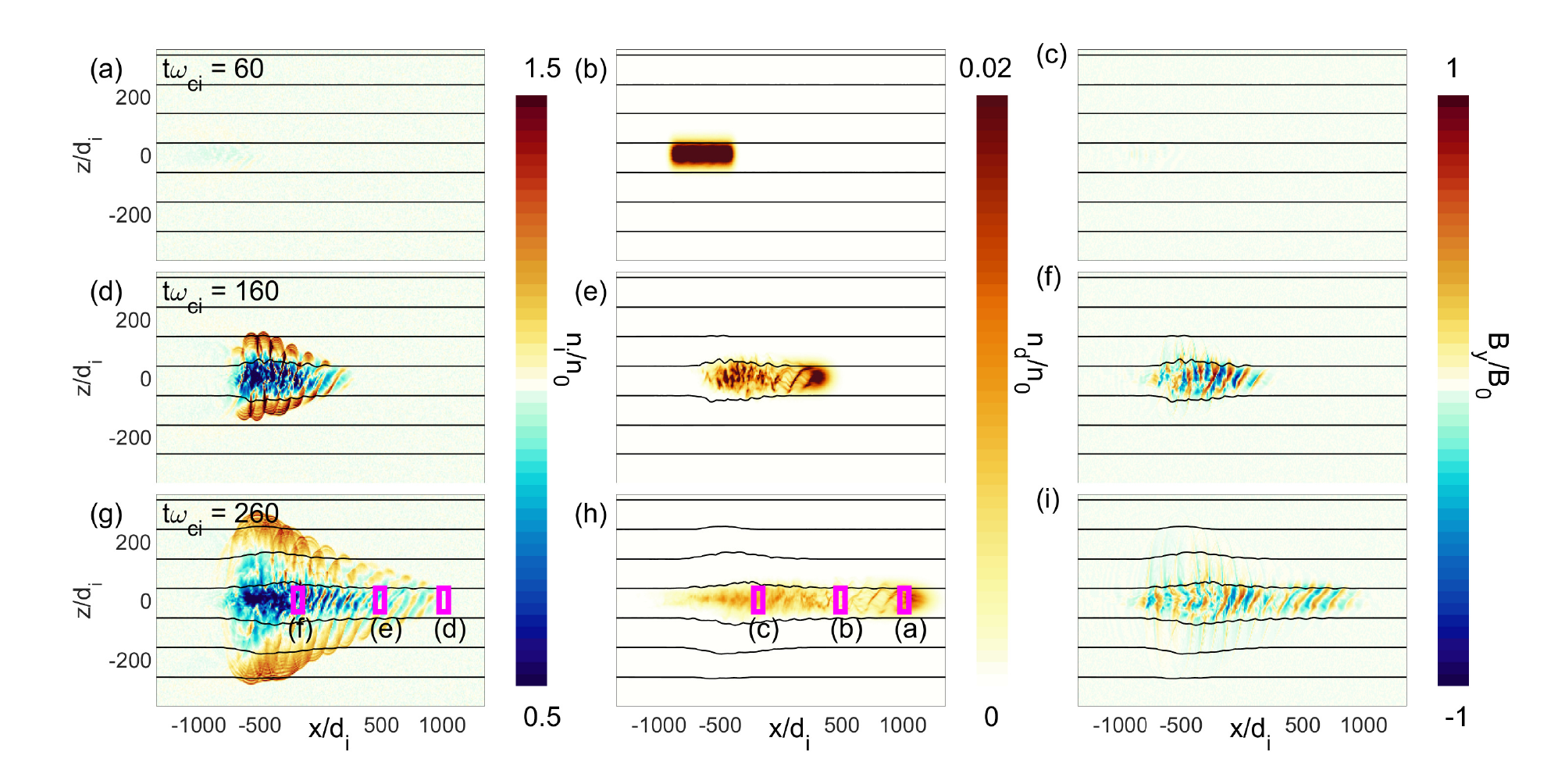}
	\caption{(a) Background density, (b) debris density, and (c) out-of-plane magnetic field component By,
each at three different times, from a 2-D hybrid simulation of a finite debris beam of length  $L_x=500~d_i$. In this case, the beam-driven instabilities saturate in a non-linear state and couple to the background plasma. The background ion density is strongly perturbed ($\delta n_i/n_i \sim 0.5$), and a large portion of the beam is slowed. The magenta boxes in (g-i) indicate where particles are sampled to compute the velocity distributions in Fig.~\ref{fig:iondist}.}
	\label{fig:finite500}
\end{figure}

The simulation of Fig.~\ref{fig:finite500} is similar to the previous calculation, but with the finite beam length $\Lambda_x=500~d_i$. In addition to increasing the total free energy in the beam, the increased beam length now supports multiple wavelengths of the unstable modes. Equivalently, the extended duration of the beam allows unstable ion beam modes to grow over a time scale of $\tau\sim\Lambda_x/v_{dx}\sim70/\omega_{ci}$. This is long enough to allow the ion-ion beam instabilities to grow to a fully saturated amplitude, with local fluctuating magnetic fields of order $\delta B/B\sim1$. The full growth of the beam instability to a non-linear state allows a strong coupling between the background ions and the beam ions. As visible in Figs.~\ref{fig:finite500}(d,g), the background ion density is strongly perturbed by the beam-driven instability. In addition, the beam ions are significantly slowed by interaction with the electromagnetic modes, resulting in a large fraction trailing behind the leading edge of the beam [see Figs.~\ref{fig:finite500}(e,h)]. This type of saturated coupling is required to generate a parallel shock front. The necessity of long enough interaction length/time, with beam ions of sufficient density to drive the non-resonant instability, places lower bounds on the total amount of ion debris required to generate a non-linear quasi-parallel shock.

\begin{figure}
	\centering
    \includegraphics[width=1\linewidth]{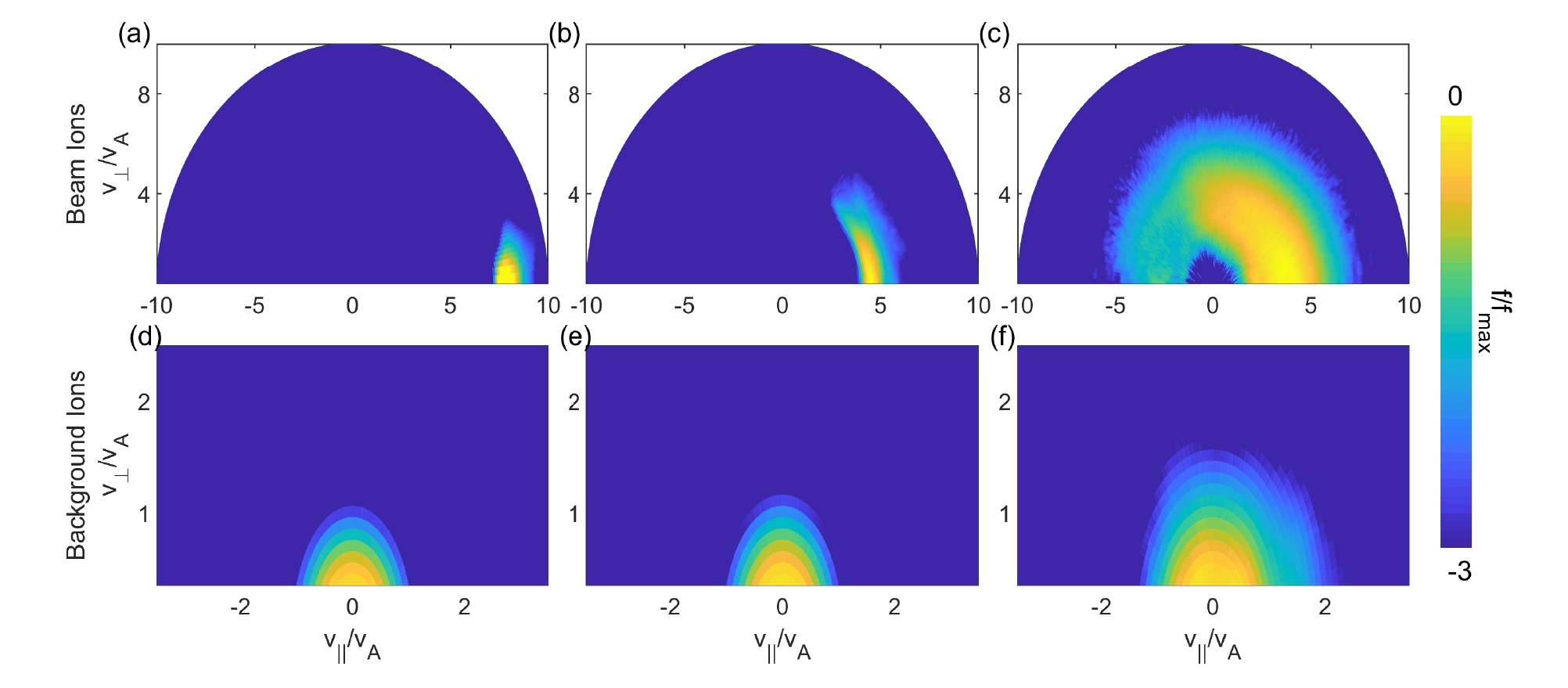}
	\caption{Ion particle distributions from the regions marked with magenta boxes in the bottom row of Fig.~\ref{fig:finite500} at time $t \times \omega_{ci}=130$ of the finite beam simulation with $\Lambda_x=500~d_i$. The beam ions show strong pitch angle scattering as well as energy diffusion (b,c), with a portion of the beam continuing at close to the initial streaming speed (a). The background ions are only mildly perturbed from their initial Maxwellian in (d,e), and they show a forward-propagating ($v_\parallel>0$) tail and evidence of heating after a longer interaction time in (f).}   
	\label{fig:iondist}
\end{figure}

The strong interaction between the beam and background ions results in substantial modifications of the beam ion velocity distribution. Reduced particle distributions as functions of $v_\parallel$ and $v_\perp$ (with respect to the local magnetic field) from the regions denoted by the three boxes marked in Fig.~\ref{fig:finite500}(g,h) are plotted in Fig.~\ref{fig:iondist}. The leading portion of fast beam ions remains close to the initial beam velocity of $M_A=7.5$ in Fig.~\ref{fig:iondist}(a). Beam ions that have undergone an intermediate reaction time with the electromagnetic instabilities show both slowing and  pitch angle scattering in Fig.~\ref{fig:iondist}(b). Meanwhile, after a longer time, the beam ions are strongly pitch-angle scattered, even into parallel velocities opposite the initial beam velocity [Fig.~\ref{fig:iondist}(c)]. The background ions are relatively unperturbed in Figs.~\ref{fig:iondist}(d,e). The background ions in Figs.~\ref{fig:iondist}(f) have had a long enough time to undergo noticeable scattering by the beam-driven instabilities, and their distribution exhibits heating, as well as a tail of background ions accelerated (by Landau damping) in the initial beam direction ($v_\parallel>0$).

Many qualitative features of the finite 2-D beam simulation reflect what is observed in the 3-D simulation of Sec.~\ref{sec:3-D}. For example, the leading edge of the fast beam drives a relatively coherent resonant instability, which does not initially couple very strongly to the background ions. While the resonant fluctuations are very similar between 2-D and 3-D, the transverse mode structure as in Eq.~\ref{eq:modelb} is different depending on the dimensionality of the system. After a long enough interaction length, stronger coupling occurs between the beam and the background. This likely includes contributions from non-resonant modes (including perhaps obliquely propagating ones), though the analysis is complicated by the strong non-uniformity that develops. As in the 3-D run, the 2-D simulations display a heating and acceleration of the background ions; see, for example, the background ion velocity $u_{ix}$ within $\sim 200d_i$ of the initial burst point in Fig.~\ref{fig:3-Dslice}. The heated background plasma expands outward leaving a lower density region, and it launches a conical wake, which is faintly visible in Fig.~\ref{fig:3-Dslice}(c) and more clear in the 2-D run of Figs.~\ref{fig:finite500}(d,g). 
\newline
\indent
Finally, we note that the perturbations in the ion background, as well as the significant slow-down in the ion beam in Fig.\ \ref{fig:finite500}, may indicate that a quasi-parallel shock is, indeed, forming in the simulation. However, the plasma is so strongly inhomogeneous that an unambiguous identification of a shock (i.e., by checking whether or not the disturbance obeys the Rankine-Hugoniot jump conditions) is not possible. Moreover, the apparent lack of a reflected ion population (which would lead the shock) in the simulation is also evidence against shock formation.

\FloatBarrier
 
\section{Conclusions}
\label{sec:disc}

Using a hybrid version of the PIC code VPIC, we explored the evolution of ion-ion beam instabilities driven by ionized debris streaming away from an explosion in a magnetized background. The modes driven by the streaming debris are modified by the 3-D geometry and gradients in the plasma parameters. Despite this, the simple linear theory [Appendix \ref{sec:1-D_beam}] agrees quite well with many properties of the beam-driven modes. We showed that the right-hand resonant (RHI) and non-resonant (NRI) ion-ion beam instabilities are present in a 3-D simulation of a point-like explosion, which represents an idealized test problem reminiscent of a multitude of systems in space, astrophysical, and laboratory settings. The leading edge of the escaping debris ions generates a modified RHI resonant electromagnetic mode with a parallel wavelength set by the beam speed and a cyclotron resonance and with a transverse mode structure that depends on the beam width and the dimensionality. For a long enough debris beam, the debris couples effectively to the background. There is evidence in the simulation for contributions from the NRI, though there are possibly also contributions from oblique modes. In any case, the coupling occurs in the presence of order-unity magnetic fluctuations driven by the resonant mode, and it is impossible to disentangle more complicated non-linear effects.  
\newline
\indent
Finally, we note that although there is some degree of debris-background coupling via RHI/NRI in the 3-D simulation, the high degree of inhomogeneity makes it difficult to discern clear evidence of quasi-parallel shocks. As we show in Section \ref{sec:finitebeam}, the finite size of the ion debris beam puts limitations on the debris-background coupling. Namely, a long enough interaction length and a sufficiently high beam density are required to drive the NRI to coupling strengths capable of generating shocks.\cite{onsager:1991} This suggests that the coupling threshold for parallel shock formation may be rather steep, and this may have implications for current laboratory efforts to observe these shocks. We leave a more thorough assessment of these questions to a future study. 

\section{Supplementary Material}
\label{sec:sup_mat}

Two videos are included online. The first video, den2-MA5-RM40-lines.avi, shows the density evolution of the 3-D point-like explosion Hybrid-VPIC simulation described in Section \ref{sec:3-D}. The second video, beam-Lx500.avi, shows the evolution of the finite ion beam in the 2-D Hybrid-VPIC simulations described in Section \ref{sec:finitebeam}.

\begin{acknowledgments}
We thank William Daughton for helpful discussions. Work performed under the auspices of the U.S.\ Department of Energy National Nuclear Security Administration under Contract No.\ 89233218CNA000001, and used resources provided by the Los Alamos National Laboratory Institutional Computing Program. This work was supported by the Laboratory Directed Research and Development program of Los Alamos National Laboratory under Project No.\ 20200334ER and by the Defense Threat Reduction Agency under projects DTRA1308134079 and DTRA1108125.
\end{acknowledgments}

\section*{Data Availability}
\label{s:data}

The data that support the findings of this study may be obtained from the corresponding author, upon reasonable request.

\appendix
\section{Linear Theory of the Right-Hand Ion-Ion Beam Instabilities} 
\label{sec:1-D_beam}

We focus here on reviewing the field-parallel resonant and non-resonant right-hand instabilities in a spatially homogeneous beam and background plasma. Both instabilities produce modes with a right-hand circular polarization (denoting the sense of rotation in time, at a fixed spatial location) in the electron rest frame (i.e., the zero current reference frame). Assuming negligible ion pressure anisotropy and that the ion beam velocity is not comparable to the background ion thermal velocity (a situation for which a separate left-hand resonant instability may exist \cite{gary:1984, gary:1991}), the linear regime (in the hybrid limit of $\omega_{ci}/\omega_{pi} = v_A/c \rightarrow 0$) for both modes is described by the dispersion relation\cite{gary:1984, winske:1984, winske:1986}
\begin{equation}     
  \hat{k}^2 \approx \sum_s\hat{\omega}_{ps}^2\xi_sZ(\xi_s^{\pm}),
\label{linear_dispersion}
\end{equation}
where $Z(x)$ is the standard plasma dispersion function, the summation is over all plasma species, $s$; and
\begin{equation}     
\begin{cases}
  \xi_s \equiv {\hat{\omega} - {\bf \hat{k}}\cdot\hat{\bf u}_s}/{\left(\sqrt{2}|{\bf \hat{k}}|\hat{v}_{ths}\right)}, \\
  \xi_s^\pm \equiv \xi_s \pm {\hat{\omega}_{cs}}/{\left(\sqrt{2}|{\bf \hat{k}}|\hat{v}_{ths}\right)},
\end{cases}
\label{xi_def}
\end{equation}
where ${\bf u}_s$ is the species drift velocity, $\hat{\omega} \equiv \hat{\omega}_r + i\hat{\gamma}$ (where $\gamma$ is the growth rate and $\omega_r$ is the real frequency), and $\hat{v}_{ths}$ is the normalized species thermal velocity. The normalizations follow as
\begin{equation}     
\begin{cases}
    \hat{\omega} \equiv \omega/\omega_{ci}, \\
    \hat{\omega}_{cs} \equiv \omega_c/\omega_{ci}, \\
    \hat{\omega}_{ps} \equiv \omega_{ps}/\omega_{pi}, \\
    \hat{k} \equiv kc/\omega_{pi} = kd_i, \\
    \hat{u} \equiv u/v_A, \\
    \hat{v}_{ths} \equiv v_{ths}/v_A= \sqrt{\beta_s/\beta_i},
\end{cases}
\label{normalizations}
\end{equation}
where $\omega_{ps} \equiv \sqrt{Z_s^2n_se^2/(m_s\epsilon_0)}$, $\beta_s = 2\mu_0n_sT_s/B_0^2$, and $i$ refers to the background ion species.
\newline
\indent
Despite the right-hand polarization of both the resonant and non-resonant modes, they can be distinguished by their helicities (i.e., the sense of rotation in space, at a fixed time).\cite{weidl:2019} The resonant mode propagates along the direction of the ambient magnetic field (i.e., $k$ > 0), but the non-resonant mode propagates in the anti-parallel direction (i.e., $k < 0$). In this case, in the electron frame, these two propagation directions -- parallel and anti-parallel -- correspond to positive and negative helicities, respectively.
\newline
\indent
As alluded to, we solve Eq.\ (\ref{linear_dispersion}) in the current-free, electron (rest) frame. For an ion beam moving through a single-ion plasma background, this frame is specified by the condition that
\begin{equation}     
Z_dn_d{\bf u}_d + Z_in_i{\bf u}_{i} - n_e{\bf u}_e = 0,
\label{electron_frame}
\end{equation}
where $d$ refers to the beam (debris) ion species.
\newline
\indent
In the linear regime, a magnetic perturbation grows as
\begin{equation}     
\partial_t|{\bf B}_{\bf k}|^2 = 2\gamma_{\bf k}|{\bf B}_{\bf k}|^2,
\label{quasi_linear}
\end{equation}
where ${\bf B}_{\bf k}$ is the k-space representation of the magnetic fluctuation/perturbation with wave-vector, ${\bf k}$, and $\gamma_{\bf k}$ is the growth rate. We may obtain ${\bf B}_{\bf k}$ from a Fast Fourier Transform (FFT) of the normalized magnetic field, $\bf{\hat{B}}$. The negative frequency components, ${\bf B}_{-\bf k}$, are the complex conjugates of the positive frequency components. Thus, we cannot directly isolate modes for which $k < 0$ via this method. However, as previously mentioned, the right-hand resonant ($k > 0$) and non-resonant ($k < 0$) modes have positive and negative helicity, respectively. Thus, a positive helicity wave (which, for the desired plasma conditions, would presumably only correspond to a resonant mode) will have strictly positive $k$, and a negative helicity (non-resonant) mode corresponds to $k < 0$.
\newline
\indent
A helicity decomposition of ${\bf B}_{\bf k}$ may be obtained by projecting ${\bf B}_{\bf k}$ onto a circular (spatial) basis\cite{weidl:2016}
\begin{equation}     
\begin{cases}
    {\bf B}_+ \equiv \left({{\bf B}_{\bf k}^y - i{\bf B}_{\bf k}^z}\right)/2, \\
    {\bf B}_- \equiv \left({{\bf B}_{\bf k}^y + i{\bf B}_{\bf k}^z}\right)/2,
\end{cases}
\label{hel_decomp}
\end{equation}
where ${\bf B}_{\pm} = 0$; for all $k < 0$. We then may apply Eq.\ (\ref{quasi_linear}) to ${\bf B}_+$ and ${\bf B}_-$ to obtain the growth rates for the positive and negative helicity perturbations, respectively.

\section{1-D Infinite Beam Hybrid-VPIC Verification Study} 
\label{sec:1-D_verify}

As a verification test of Hybrid-VPIC, we compare the solution of Eq.\ (\ref{linear_dispersion}) -- the gray (``dash-dot'') curve -- to the results obtained by a 1-D hybrid simulation. The linear theory, as developed above, formally applies to a homogeneous medium with an infinite beam exactly parallel to the magnetic field, hence only a 1-D simulation (in this case, with periodic boundary conditions) is required for comparisons. Note that the negative helicity modes are projected on negative values of k in the figure. Here, $n_d/n_e = 0.05$, $M_A \equiv u_d/v_A = 10$, $m_i/m_p = m_d/m_p  = Z_i = Z_d = 1$ (where $m_p$ is the proton mass), and $T_i = 10T_d = T_e$. 
\begin{figure}[htbp]
    \centering
    \begin{subfigure}[b]{0.35\textwidth}
        \includegraphics[width=1.0\textwidth]{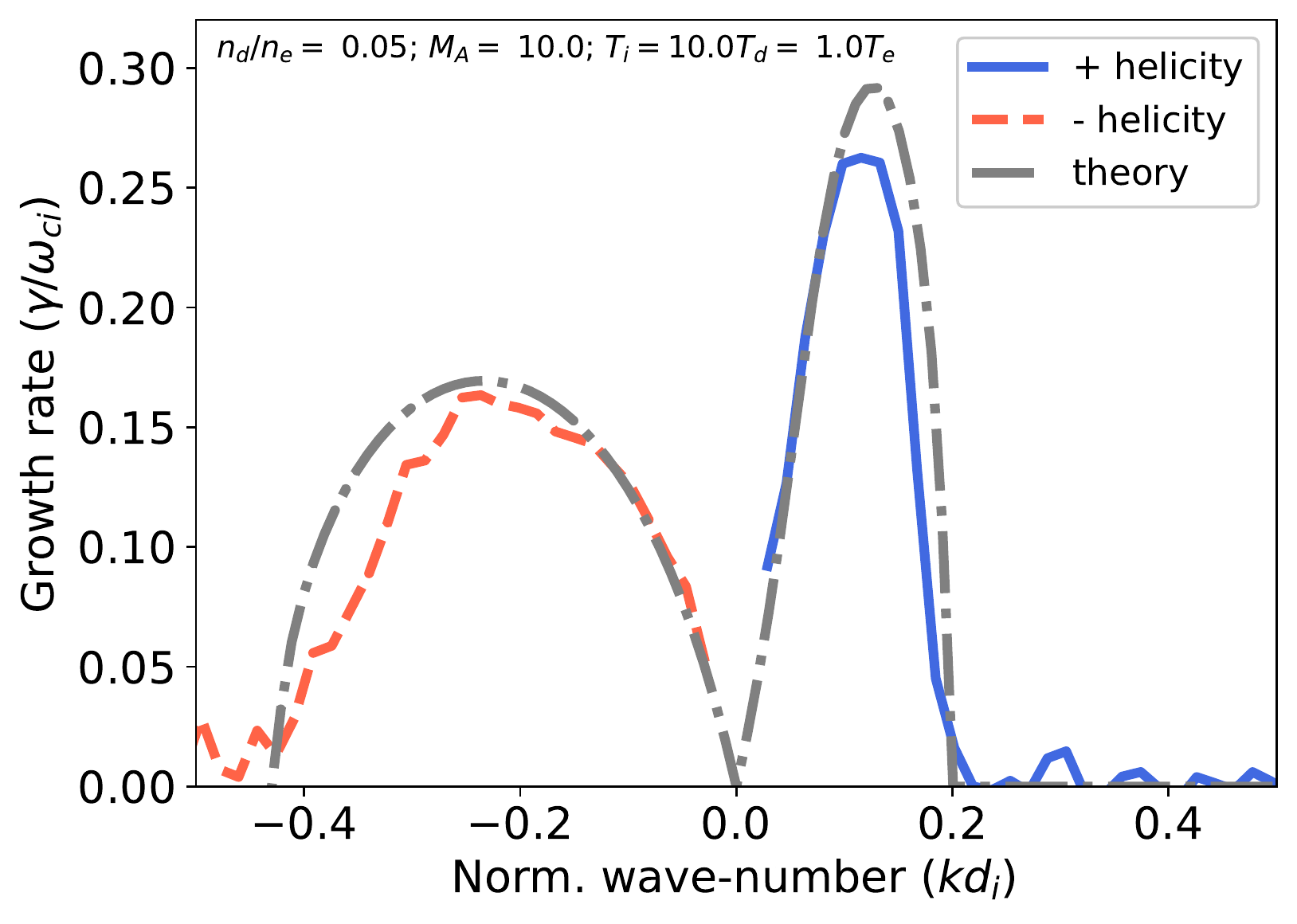}
        \caption{Growth rate vs.\ wave-number.}
        \label{1-D_gamma_vs_k}
    \end{subfigure}
    \begin{subfigure}[b]{0.35\textwidth}
        \includegraphics[width=1.0\textwidth]{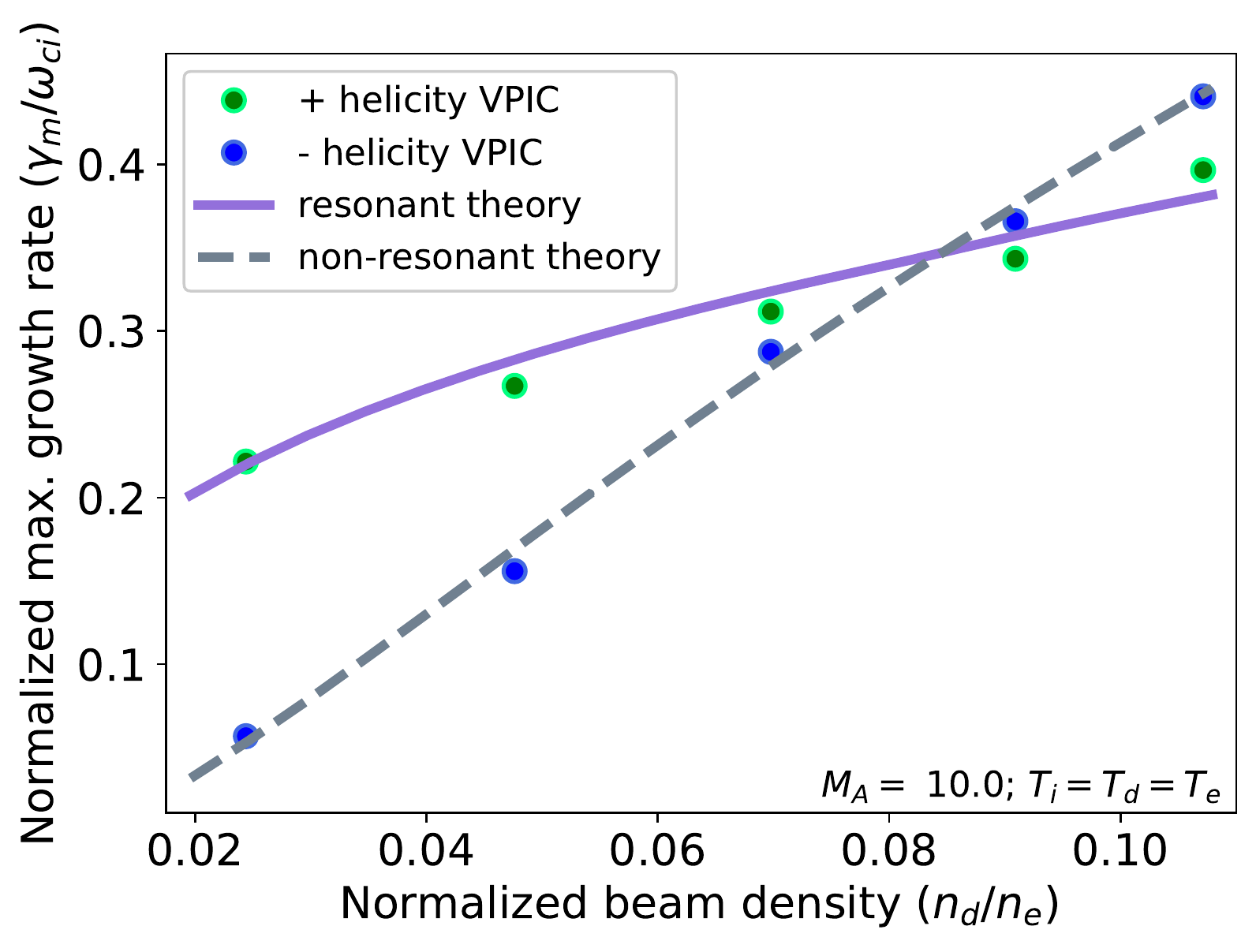}
        \caption{Growth rate vs.\ the beam density.}
        \label{growth_vs_f}
    \end{subfigure}
    \caption{VPIC/Theory comparisons in a 1-D idealized plasma.}
\end{figure}
We obtain a growth rate from the VPIC data by isolating the linear segment of $|{\bf B}_{\bf k}|^2$ vs.\ time. Then, in accord with Eq.\ (\ref{quasi_linear}), one-half of the slope of this segment should be equal to the growth rate (for a given $k$). We see decent agreement between the linear theory prediction and the numerical growth rate, $\hat{\gamma}$ vs.\ the wave-number, $\hat{k}$. Finally, Fig.\ \ref{growth_vs_f} shows the growth rate -- for both the resonant and non-resonant modes -- vs.\ the normalized beam density, $n_d/n_e$. In this theory/code comparison, $M_A = 10$ and $T_i = T_d = T_e$. Once again, we see good agreement between Hybrid-VPIC and theory.

\FloatBarrier

\bibliography{1140799_1_art_file_18815614_r3yvh9.bib}

\end{document}